\documentstyle[aps,prl]{revtex}

\title{Separability of very noisy mixed states and implications for NMR 
quantum computing}

\author{S.~L.~Braunstein,$^1$ C.~M.~Caves,$^2$ R.~Jozsa,$^3$ N.~Linden,$^4$ 
S.~Popescu,$^{4,5}$ and R.~Schack$^{2,6}$}

\address{
$^1$SEECS, University of Wales, Bangor LL57 1UT, UK\\
$^2$Center for Advanced Studies, Department of Physics and Astronomy,\\ 
University of New Mexico, Albuquerque, New Mexico 87131-1156, USA \\
$^3$School of Mathematics and Statistics, University of Plymouth, 
Devon PL4 8AA, UK \\
$^4$Isaac Newton Institute for Mathematical Sciences, 
Cambridge, CB3 0EH, UK\\
$^5$BRIMS, Hewlett-Packard Laboratories, Stoke Gifford, 
Bristol BS12 6QZ, UK\\
$^6$Department of Mathematics, Royal Holloway, University of London,
Egham, Surrey TW20 0EX, UK
}

\date{1998 December 2}

\begin{document}

\twocolumn

\draft

\maketitle

\begin{abstract}
We give a constructive proof that all mixed states of $N$ qubits in a 
sufficiently small neighborhood of the maximally mixed state are separable. 
The construction provides an explicit representation of any such state 
as a mixture of product states. We give upper and lower bounds on the 
size of the neighborhood, which show that its extent decreases exponentially 
with the number of qubits. We also discuss the implications of the bounds 
for NMR quantum computing. 
\end{abstract}


In this Letter we investigate the structure of the space of density matrices
of $N$ spin-1/2 particles (qubits).  In particular, we consider density 
matrices that are close to the maximally mixed density matrix and ask 
whether or not they are separable, i.e., whether they can be written as
mixtures of direct products.  One might imagine that the issue is 
straightforward, in that the maximally mixed state seems to be very far 
from the boundary between separable and nonseparable states.  It might be 
the case, however, that the maximally mixed density matrix is surrounded 
by separable matrices, but that these separable density matrices lie in a 
low-dimensional subspace within the space of all density matrices.  By 
leaving this subspace, even infinitesimally, one could reach entangled 
density matrices.  

In \cite{Zyczkowski} this problem is addressed by an existence proof; namely, 
it is shown that there exists a sufficiently small neighborhood of the 
maximally mixed density matrix inside which all density matrices are 
separable. 
In \cite{Vidal98}, a lower bound on the size of the neighborhood is given.   
Here we go further by giving a constructive proof that provides
an explicit representation of any state sufficiently close to the maximally 
mixed one as a mixture of product states. We give an upper and a much improved
lower bound
on the size of the neighborhood, which show that it
decreases exponentially with the number of qubits.
 
Our results have immediate implications for present research that makes
use of high-temperature nuclear magnetic resonance (NMR) for quantum 
information processing and quantum computation 
\cite{Cory97a,Cory97b,Gershenfeld97,Chuang97,Chuang98a,Chuang98b,Jones98a,%
Jones98b,Cory98,Laflamme98,LBJ,Nielsen98}.  
Since the first proposals to use NMR for quantum computation, there has 
been surprise about the apparent ability to perform quantum computations 
in room-temperature thermal ensembles.  It has been a puzzle how these
thermal states, which are very close to the maximally mixed state, could 
correspond to truly entangled states \cite{DiVin}.  The bounds we calculate 
show that {\it all states so far used in NMR to simulate quantum computations 
or other quantum-information protocols are separable.}  This is not to say 
that NMR techniques are incapable of producing entangled states, in principle.  
Increasing the number of correlated spins might eventually lead to nonseparable 
states, but this question is left open by the bounds derived in this paper.

We consider arbitrary density matrices for $N$ qubits, written as
\begin{eqnarray}
\rho_\epsilon = (1-\epsilon) M_d + \epsilon \rho_1
\;,
\label{rhoepsilon}
\end{eqnarray}
where $d=2^N$ is the Hilbert-space dimension for $N$ qubits, $M_d=1_d/d$
is the maximally mixed density matrix ($1_d$ is the identity matrix in $d$ 
dimensions), and $\rho_1$ is an arbitrary density matrix.  Any density matrix
can be written in the form~(\ref{rhoepsilon}).  We show that for 
$\epsilon$ sufficiently small, all density matrices of the 
form~(\ref{rhoepsilon}) are separable.  We define two kinds of 
representations of $\rho_\epsilon$ in terms of product states, which 
provide candidates for ensemble decompositions of $\rho_\epsilon$ as a 
mixture of product states. By considering these candidate decompositions, 
we derive an explicit lower bound on the size of the neighborhood of 
separable states.  We conclude by establishing an explicit upper bound 
on the size of the neighborhood.

Our approach is to represent an arbitrary density matrix in an overcomplete 
matrix basis, each basis element of which is a pure direct-product density 
matrix.  If all the coefficients of a density matrix in this representation 
are nonnegative, the coefficients can be considered to represent probabilities, 
and the density matrix is separable, as it is then a mixture of direct 
products.  

All of our representations arise ultimately from expanding a density matrix
for $N$ qubits in terms of direct products of Pauli matrices:
\begin{equation}
\rho=
{1\over2^N}
c_{{\alpha_1}\ldots{\alpha_N}}
\sigma_{\alpha_1}\otimes\cdots\otimes\sigma_{\alpha_N}
\;.
\label{Paulirho}
\end{equation} 
Here and throughout we sum over repeated indices: Greek indices run over 
the values 0,1,2,3, and Latin indices take on the values 1,2,3.  The matrix 
$\sigma_0=1_2$ is the two-dimensional identity matrix, and the matrices 
$\sigma_i$, $i=1,2,3$, are the Pauli matrices.  The (real) expansion 
coefficients in Eq.~(\ref{Paulirho}) are given by
\begin{equation}
c_{{\alpha_1}\ldots{\alpha_N}}
={\rm tr}\bigl(\rho\,
\sigma_{\alpha_1}\otimes\cdots\otimes\sigma_{\alpha_N}
\bigr)
\;.
\label{calpha}
\end{equation}
Normalization requires that $c_{0\ldots0}=1$.  Since the eigenvalues of 
the Pauli matrices are $\pm1$, the expansion coefficients satisfy
\begin{equation}
-1\le
c_{{\alpha_1}\ldots{\alpha_N}}
\le1
\label{cbound}
\;.
\end{equation}

To be concrete, we consider first the case of two qubits.  For each qubit
we introduce six pure density matrices, $P_i\equiv{1\over 2} (1_2 + \sigma_i)$ 
and  $\overline P_i\equiv{1\over 2} (1_2 - \sigma_i)$.  A convenient 
discrete overcomplete basis for discussing separability consists of the 
36 direct-product projectors, each of which is a pure direct-product
density matrix: $P_i\otimes P_j$, $P_i\otimes \overline P_j$, 
$\overline P_i\otimes P_j$, $\overline P_i\otimes \overline P_j$.
%\begin{equation}
%P_i\otimes P_j
%\;,\quad 
%P_i\otimes \overline P_j 
%\;,\quad 
%\overline P_i\otimes P_j 
%\;,\quad 
%\overline P_i\otimes \overline P_j
%\;. 
%\label{discrete_basis_2}
%\end{equation}
Any density matrix of two qubits can be expanded in this basis, but since 
the basis is overcomplete, the representation is not unique.  We make a 
specific choice, as follows.  Noting that $\sigma_i=P_i-\overline P_i$ and 
$1_2=P_i+\overline P_i$, we can write $1_2=\omega_i (P_i + \overline P_i)$,
where $\omega_i=1/3$, $i=1,2,3$.  With these results we can convert the
Pauli representation~(\ref{Paulirho}) into the form
\begin{eqnarray}
\rho &=& 
{1\over4}\bigl[\bigl( 
\omega_i\omega_j +c_{i0}\omega_j + \omega_ic_{0j} + c_{ij}
\bigr) 
P_i\otimes P_j\nonumber\\
&\mbox{}&
\hphantom{{1\over4}} 
+ 
\bigl( 
\omega_i\omega_j -c_{i0}\omega_j + \omega_ic_{0j} - c_{ij}
\bigr) 
\overline P_i\otimes P_j\nonumber\\
&\mbox{}&
\hphantom{{1\over4}} 
+ 
\bigl( 
\omega_i\omega_j +c_{i0}\omega_j - \omega_ic_{0j}- c_{ij}
\bigr) 
P_i\otimes \overline P_j\nonumber\\
&\mbox{}&
\hphantom{{1\over4}} 
+ 
\bigl( 
\omega_i\omega_j -c_{i0}\omega_j - \omega_ic_{0j}+ c_{ij}
\bigr) 
\overline P_i\otimes \overline P_j 
\bigr]\;. 
\label{rho_representation}
\end{eqnarray}
If the coefficient of each of the 36 basis elements is nonnegative, the 
density matrix is separable.  We note that when the maximally mixed 
density matrix for two qubits, $M_4={1\over 4}1_2 \otimes 1_2$, is
represented as in Eq.~(\ref{rho_representation}), the coefficient of 
each of the basis matrices is $1/36$.

Consider now an arbitrary entangled (nonseparable) density matrix $\rho_1$.  
Since $\rho_1$ is entangled, at least one of the coefficients in the 
representation of $\rho_1$ in the form~(\ref{rho_representation}) is 
negative.  Suppose now that $\rho_1$ is mixed with the maximally mixed 
density matrix $M_4$ as in Eq.~(\ref{rhoepsilon}), i.e., 
$\rho_\epsilon = (1-\epsilon)M_4 + \epsilon\rho_1$.  Although some of the 
coefficients of $\rho_1$ are negative, all of the coefficients of $M_4$ 
are strictly positive.  Hence, for $\epsilon$ small enough, all the 
coefficients of $\rho_\epsilon$ are nonnegative, making $\rho_\epsilon$ 
separable.  Thus {\it all\/} density matrices in a sufficiently small 
neighborhood of the maximally mixed density matrix are separable.  

Furthermore, we can find an explicit bound on $\epsilon$ such that
$\rho_\epsilon$ is separable for any $\rho_1$.  To find a bound, we 
use Eq.~(\ref{cbound}) to bound the coefficients of the basis matrices
in a representation of $\rho_1$ of the form~(\ref{rho_representation}).  
The minimum value of any of the coefficients is $(1/4)(1/9-1/3-1/3-1)=-14/36$.  
Thus all the coefficients of the density matrix $\rho_\epsilon$ in the 
discrete overcomplete basis are nonnegative if
$(1-\epsilon)/36-14\epsilon/36\geq0$, i.e., if $\epsilon\le1/15$.  For 
$\epsilon\le1/15$, the representation~(\ref{rho_representation}) is an 
explicit decomposition of $\rho_\epsilon$ as a mixture of direct products.  

A similar analysis can be carried out for any number of qubits.  Starting 
from the Pauli representation~(\ref{Paulirho}), we introduce a discrete
product basis, like that for two qubits, and define a representation 
analogous to that in Eq.~(\ref{rho_representation}).  Using Eq.~(\ref{cbound})
to limit the size of the coefficients in this representation, we find an 
asymptotic lower bound on the size of the neighborhood of separable density 
matrices that is of order $\epsilon\sim1/4^N$ for $N$ qubits.

One particularly interesting example is the GHZ state 
\cite{Laflamme98,Greenberger90}, a state for three qubits whose density 
matrix is 
\begin{eqnarray}
&\mbox{}&
\rho_{\rm GHZ}
\nonumber\\
&\mbox{}&\hphantom{\rho}
={1\over 2}
\bigl(|111\rangle+|222\rangle\bigr)\bigl(\langle111|+\langle222|\bigr)
\nonumber\\
&\mbox{}&\hphantom{\rho}
={1\over 8}\Bigl( 
1_2\otimes 1_2\otimes 1_2 
+1_2\otimes\sigma_3\otimes\sigma_3
+\sigma_3\otimes 1_2\otimes \sigma_3
\nonumber\\
&\mbox{}&\hphantom{\rho={1\over8}\Bigl(}
+\sigma_3\otimes\sigma_3\otimes1_2
+\sigma_1 \otimes \sigma_1 \otimes \sigma_1 
-\sigma_1 \otimes \sigma_2 \otimes \sigma_2\quad
\nonumber\\
&\mbox{}&\hphantom{\rho={1\over8}\Bigl(}
-\sigma_2 \otimes \sigma_1 \otimes \sigma_2
-\sigma_2 \otimes \sigma_2 \otimes \sigma_1
\Bigr)
\;.
\label{GHZstate}
\end{eqnarray}
We now express the maximally mixed density matrix, $M_8$, and the GHZ 
density matrix in terms of an overcomplete set of $6^3=216$ basis 
matrices analogous to the 2-qubit matrices introduced above.  
We find that $M_8$ has coefficient $1/216$ for all the basis elements 
and that the smallest coefficient for $\rho_{\rm GHZ}$ in this basis 
is $-(1/8)(26/27)=-26/216$.  Thus for $\epsilon\le1/27$, the state
\begin{equation}
\rho_\epsilon=(1-\epsilon)M_8 + \epsilon\rho_{\rm GHZ} 
\label{GHZmixture}
\end{equation}
is separable.  We return to the GHZ example below.

We have also considered another overcomplete basis for the space of density 
matrices, a basis labeled by continuous parameters.  An arbitrary density 
matrix for $N$ qubits can be represented as
\begin{equation}
\rho=\int
d\Omega_1\cdots d\Omega_N\,
w({\vec n}_1,\ldots,{\vec n}_N)
P_{{\vec n}_1}\otimes\cdots\otimes P_{{\vec n}_N}
\;,
\label{cts_expn}
\end{equation}
where the integral runs over $N$ Bloch spheres and where 
$P_{\vec n}\equiv{1\over2}(1_2+\vec n\cdot\vec\sigma)$
is the projector onto the pure state located at unit vector $\vec n$ on 
the Bloch sphere.  The representation (\ref{cts_expn}) is by no means 
unique.  In a spherical-harmonic expansion of 
$w({\vec n}_1,\ldots,{\vec n}_N)$, the density matrix determines only 
the $l=0$ and $l=1$ parts; the higher-order spherical-harmonic content 
corresponds to the freedom in representing $\rho$ as a sum of one-dimensional 
product projectors.  A separable density matrix is one for which there 
is an expansion such that $w({\vec n}_1,\ldots,{\vec n}_N)$ is everywhere 
nonnegative. 

We can generate a candidate for a separable ensemble decomposition of
$\rho$ by considering the unique representation of the form~(\ref{cts_expn}) 
such that $w({\vec n}_1,\ldots,{\vec n}_N) $ has only $l=0$ and $l=1$ 
components.  We can obtain this unique representation by noting that
\begin{equation}
{1\over2}\sigma_\alpha=
{3\over4\pi}\int d\Omega\,n_{\alpha}P_{\vec n}
\;,
\end{equation}
where $n_0\equiv1/3$.  Inserting this result into the Pauli-matrix
expansion~(\ref{Paulirho}) and using Eq.~(\ref{calpha}) gives 
\begin{eqnarray}
&\mbox{}&
w({\vec n}_1,\ldots,{\vec n}_N)
\nonumber\\
&\mbox{}&\hphantom{i}
=
\left({3\over4\pi}\right)^{\!N}\!
c_{\alpha_1\ldots\alpha_N}
(n_1)_{\alpha_1}\cdots (n_N)_{\alpha_N}
\nonumber\\
&\mbox{}&\hphantom{i}
=
{1\over(4\pi)^N}
{\rm tr}
\Bigl(
\rho
(1_2+3\vec n_1\cdot\vec\sigma)
\otimes\cdots\otimes
(1_2+3\vec n_N\cdot\vec\sigma)
\Bigr)
\;.
\nonumber\\
\label{probability}
\end{eqnarray}
The maximally mixed density matrix, $M_{2^N}$, has $w=(1/4\pi)^N$.

Let us concentrate on the operator product in the last form of 
Eq.~(\ref{probability}).  Each operator in the product has eigenvalues 
4 and $-2$.  Thus the most negative eigenvalue of the operator product 
is $4^{N-1}(-2)=-2^{2N-1}$, which implies that
\begin{equation}
w({\vec n}_1,\ldots,{\vec n}_N)
\ge
-{2^{2N-1}\over(4\pi)^N}
\;.
\end{equation}

Consider now the density matrix~(\ref{rhoepsilon}).  Its candidate
ensemble probability satisfies
\begin{eqnarray}
w_\epsilon({\vec n}_1,\ldots,{\vec n}_N)&=&
{1-\epsilon\over(4\pi)^N}+
\epsilon w_1({\vec n}_1,\ldots,{\vec n}_N)
\nonumber\\
&\ge&
{1-\epsilon(1+2^{2N-1})\over(4\pi)^N}
\;.
\end{eqnarray}
Therefore $\rho_\epsilon$ is separable if
\begin{equation}
\epsilon\le{1\over1+2^{2N-1}}
\mathop{\sim}_{N\rightarrow\infty}
{2\over4^N}
\;.
\end{equation}
We see again that all density matrices in the neighborhood
of the maximally mixed density matrix are separable, and we obtain
a lower bound on the size of the separable neighborhood, which
for large $N$ is much better than the bound,
$\epsilon\le(1+2^{N-1})^{-(N-1)}$, given in \cite{Vidal98}.

It is instructive to return to the GHZ state~(\ref{GHZstate}) and to note 
that Eq.~(\ref{probability}) gives
\begin{eqnarray}
w_{\rm GHZ}(\vec n_1,\vec n_2,\vec n_3)
&=&{1\over(4\pi)^3}
\bigl[1+9(c_1c_2+c_2c_3+c_1c_3)
\nonumber\\
&\mbox{}&\hphantom{{1\over(4\pi)}}
+27s_1s_2s_3\cos(\varphi_1+\varphi_2+\varphi_3)
\bigr]
\nonumber\\
&\ge&
-{26\over(4\pi)^3}
\;.
\label{wGHZ}
\end{eqnarray}
Here $c_j\equiv\cos\theta_j$ and $s_j\equiv\sin\theta_j$, and the 
minimum value occurs at $\theta_1=\theta_2=\theta_3=\pi/2$ and 
$\varphi_1+\varphi_2+\varphi_3=\pi$.  Equation~(\ref{wGHZ}) shows that 
the mixed state~(\ref{GHZmixture}) is separable if $\epsilon\le1/27$, 
the same bound obtained above.  

This bound is not optimal.  To find a bound for a particular state, 
such as the GHZ state, one should expand the state in terms of a 
tailor-made set of direct products, instead of a general-purpose 
set.  The continuous set in Eq.~(\ref{cts_expn}) provides a starting 
point for developing a more efficient representation (treated in an
upcoming publication) that uses a linearly independent set of $4^N$ 
direct products of the form 
$P_{{\vec n}_1}\otimes\cdots\otimes P_{{\vec n}_N}$.  The explicit 
form of this representation is
\begin{equation}
\rho=
\sum_{\vec n_1,\ldots,\vec n_N}
w({\vec n}_1,\ldots,{\vec n}_N)
P_{{\vec n}_1}\otimes\cdots\otimes P_{{\vec n}_N}
\;,
\end{equation}
where $w({\vec n}_1,\ldots,{\vec n}_N)$ is given by Eq.~(\ref{probability}),
and the sum runs over Bloch vectors that lie at the vertices of tetrahedra.  
Using a representation of this sort matched to the GHZ state, one can 
improve the bound for separability of the state~(\ref{GHZmixture}) to 
$\epsilon\le1/(3+6\sqrt2)\simeq1/11.5$. 

In this Letter we have been using $\epsilon$ to characterize how close
a density matrix of the form~(\ref{rhoepsilon}) is to the maximally
mixed density matrix, $M_d$.  An alternative distance measure, defined
by $\delta\equiv\sqrt{{\rm tr}\bigl((\rho-M_d)^2\bigr)}$, leads to 
similar overall conclusions, for using the representation of $\rho$ in 
Eq.~(\ref{probability}), one can show that all states with 
$\delta\le1/(2\sqrt5)^N$ are separable.  

Up to this point we have been thinking of the number of qubits as being 
fixed, and we have investigated the boundary between separability and 
nonseparability as the amount of noise, specified by $\epsilon$, changes.  
We now shift gears, thinking of the qubits as particles with spin and 
asking what happens as the number of particles or their dimension changes, 
while $\epsilon$ is held fixed.  In general, as we go to more particles 
or higher spins, we find that we can tolerate more mixing with the 
maximally mixed state and still have states that are not separable.  
In other words, for a given $\epsilon$, we can always find states of 
sufficiently large numbers of particles or sufficiently high spin for 
which $\rho_\epsilon$ is nonseparable.  We translate this result into 
an upper bound on the size of the separable neighborhood around the 
maximally mixed state.  

Consider now two spin-$(d-1)/2$ particles, each living in a 
$d$-dimensional Hilbert space.  What we have in mind is that 
each of these particles is an aggregate of $N/2$ spin-1/2 particles 
(qubits), in which case $d=2^{N/2}$.  We consider a specific joint 
density matrix of the two particles, 
\begin{equation}
\rho_{\epsilon}=(1-\epsilon) M_{d^2} + 
\epsilon|\psi\rangle\langle\psi|
\;.
\label{mixed_plus_entangled}
\end{equation}
where $|\psi\rangle$ is a maximally entangled state of the two particles, 
\begin{equation}
|\psi\rangle=
{1\over\sqrt d}
\bigl(
|1\rangle|1\rangle+|2\rangle|2\rangle+...+|d\rangle|d\rangle
\bigr)
\;.
\end{equation}

Now project each particle onto the subspace spanned by $|1\rangle$ and 
$|2\rangle$.  The state after projection is
\begin{eqnarray}
\tilde\rho&=& 
{1\over A}
\biggl({1-\epsilon\over d^2}1_{4}
+{\epsilon\over d}
\Bigl(
|1\rangle|1\rangle+|2\rangle|2\rangle
\Bigr)
\Bigl
(\langle 1|\langle 1| + \langle 2|\langle 2|
\Bigr)
\biggr)
\nonumber\\
&=&  
(1-\epsilon')M_4+\epsilon'|\phi\rangle\langle\phi|
\;,
\label{entangled}
\end{eqnarray}
where $A=(4/d^2)[1+\epsilon(d/2-1)]$ is the normalization factor, 
\begin{equation}
|\phi\rangle={1\over\sqrt2}
\bigl(|1\rangle|1\rangle+|2\rangle|2\rangle\bigr)
\end{equation}
is a maximally entangled state of two qubits, and 
\begin{equation}
\epsilon'=
{2\epsilon/d\over A}=
{\epsilon d/2\over1+\epsilon(d/2-1)}
\;.
\end{equation}

The projected state $\tilde\rho$ is a Werner state \cite{Werner}, a 
mixture of the maximally mixed state for two qubits, $M_4$, and the
maximally entangled state $|\phi\rangle$.  The proportion $\epsilon'$ 
of maximally entangled state increases linearly with $d$.  Thus, 
as $d$ increases for fixed $\epsilon$, there is a critical dimension 
beyond which $\tilde\rho$ becomes entangled.  Indeed, the Werner state 
is nonseparable for $\epsilon'>1/3$ \cite{Werner,Wernerproof}, which is 
equivalent to $d>\epsilon^{-1}-1$.  Moreover, since the local 
projections on the two particles cannot create entanglement from a 
separable state, we can conclude that the state~(\ref{mixed_plus_entangled}) 
of $N$ qubits is nonseparable under the same conditions, i.e., if
\begin{equation}
\epsilon>
{1\over1+d}=
{1\over1+2^{N/2}}
\;.
\end{equation}
This result establishes an {\it upper\/} bound, scaling like $2^{-N/2}$, on 
the size of the separable neighborhood around the maximally mixed state.  

Our results have implications for attempts to use high-temperature 
NMR techniques to perform quantum computations or other 
quantum-information-processing tasks.  They imply that NMR experiments 
performed to date have not produced genuinely entangled density matrices.  
This is because in current experiments, the parameter $\epsilon$, which 
measures the deviation from the maximally mixed state, has a value 
$\sim10^{-5}$, much smaller than the lower bounds we have found for the 
radius of the separable neighborhood of the maximally mixed state, for 
the cases of two or three spins used in these experiments.  

Present high-temperature NMR techniques, based on synthesizing a 
pseudopure state in the deviation density matrix, imply that $\epsilon$ 
scales like $N/2^N$ as the number of qubits increases at constant
temperature \cite{Cory97a,Gershenfeld97}.  With this scaling, 
the state $\rho_\epsilon$
leaves the region where our lower bound implies that all states are 
separable at about 14 qubits, but it never enters the region where our 
upper bound guarantees that there are entangled states.  Thus, it is
unclear whether present NMR techniques can produce entangled states.  
Different techniques might lead to a more favorable 
scaling behavior for $\epsilon$ \cite{Vazirani}.

The results in this Letter suggest that current NMR experiments 
should be considered as simulations of quantum computations rather than 
true quantum computations, since no entanglement appears in the physical 
states at any stage of the process \cite{ekertjozsa}.  We stress, however, 
that we have not given a proof of this conclusion, since we would need to
analyze further the power of general unitary operations in their action
on separable states.  Much more needs to be understood about what it means 
for a computation to be a ``quantum'' computation.

SLB, RJ, and RS are supported by the UK Engineering and Physical
Sciences Research Council.  RJ is supported in part by the European TMR
Research Network ERB-FMRX-CT96-0087.  CMC is supported in part by the 
US Office of Naval Research N00014-93-1-0116.  SLB, RJ, and RS acknowledge 
the support and hospitality of the Workshop on Quantum Information,
Decoherence and Chaos held on Heron Island, Queensland, in September
1998, where the issues in this work were raised, and are grateful to
T.~F. Havel and R. Laflamme for discussions about NMR computing at
that workshop.

\end{document}